\documentclass[prd,aps,twocolumn,showpacs,preprintnumbers,amsmath,amssymb,nofootinbib]{revtex4}
\usepackage[dvips]{graphicx}
\usepackage{epsf}
\usepackage{amsmath}
\usepackage{amssymb}
\usepackage{tensor}
\usepackage{graphicx}
\usepackage{dcolumn}
\usepackage{bm}
\usepackage{color}
\usepackage[colorlinks]{hyperref}
\begin{document}
\title{Revisiting f(R) gravity models that reproduce $\Lambda$CDM expansion}

\author{Jian-hua He$^{1}$\email{jianhua.he@brera.inaf.it}, Bin Wang$^{2}$}

\affiliation{$^{1}$ INAF-Osservatorio Astronomico
di Brera, Via Emilio Bianchi, 46, I-23807, Merate
(LC), Italy}
\affiliation{$^{2}$ INPAC and
Department of Physics, Shanghai Jiao Tong
University, Shanghai 200240, China}


\begin{abstract}
We reconstruct an $f(R)$ gravity model that gives rise to the particular $\Lambda$CDM background evolution of the universe. We find well-defined, real-valued analytical forms for the $f(R)$ model to describe the universe both in the early epoch from the radiation to matter dominated eras and the late time acceleration period.  We further examine the viability of the derived $f(R)$ model and find that it is viable to describe the evolution of the universe in the past and there does not exist the future singularity in the Lagrangian.
\end{abstract}

\maketitle

\section{Introduction}
Cosmological observations from supernovae\cite{1}, BAO~\cite{BAOmeasurement}and CMB \cite{WMAP} indicate that our universe is undergoing a phase of accelerated expansion. Understanding the nature of the cosmic acceleration is one of the biggest questions in modern physics. This acceleration is believed to be driven by a so called dark energy (DE) in the framework of Einstein's general relativity. The simplest explanation of such DE is the cosmological constant. However, the measured value of the cosmological constant is far below the prediction of any sensible quantum field theories and furthermore the cosmological constant leads inevitably to the coincidence problem, namely why the energy densities of matter and the vacuum are of the same order today(see \cite{sean} for review).

Alternatively, the acceleration can be explained by modifying the gravity theory. The theory of general relativity might not be ultimately correct on cosmological scales. One of the simplest attempts is called $f(R)$ gravity, in which the
scalar curvature in the Lagrangian density of
Einstein's gravity is replaced by an arbitrary
function of $R$. However the complexity of the field equations makes it difficult to obtain a viable $f(R)$ model to satisfy both cosmological and local
gravity constraints\cite{fr}. Recently, there appeared a useful approach to reconstruct the $f(R)$ model by inverting the observed expansion history of the universe to deduce what class of $f(R)$ theories give rise to the particular cosmological evolution  \cite{Song}\cite{Pogosian}\cite{Lombriser}\cite{dobado}\cite{Dunsby}\cite{Nojiri}\cite{solution}. Some analytical forms for $f(R)$ gravity that admit the $\Lambda$CDM expansion history in the background spacetime were constructed \cite{dobado} \cite{Dunsby}\cite{Nojiri}. However, it was argued that only a simple real-valued expression of $f(R)$ model in the Lagrangian could admit an exact $\Lambda$CDM expansion history\cite{Dunsby}.

In this paper, we will further study this problem. We will perform a number of explicit reconstructions which lead to a number of interesting results. We will show that we can derive a well-defined real-valued analytical $f(R)$ in terms of the hypergeometric functions to admit an exact $\Lambda$CDM expansion history. We will explicitly show that the $f(R)$ gravity not only can admit an exact $\Lambda$CDM expansion in the recent epoch of the universe but also can admit an exact $\Lambda$CDM expansion in the early time of the universe. We will also discuss the physical boundary conditions for these constructions.

This paper is organized as follows: In section~\ref{RM}, we review the background dynamics of the universe in the $f(R)$ gravity and present the well-defined, real-valued analytical $f(R)$ forms that can exactly reproduce the same background expansion as that of the $\Lambda$CDM model from the radiation dominated epoch to the matter dominated epoch.
In section~\ref{MA}, we present the explicit form for a $f(R)$ model that can mimic the evolution of the universe from the matter dominated epoch to the late time acceleration. We will also discuss the physical boundary conditions and viability for these models. In section~\ref{conclusions}, we will summarize and conclude this work.
\section{The transition from the radiation dominated epoch to the matter dominated era \label{RM}}
We work with the 4-dimensional action in the $f(R)$
gravity\cite{frreview}
\begin{equation}
S=\frac{1}{2\kappa^2}\int d^4x\sqrt{-g}f(R)+\int
d^4x\mathcal{L}^{(m)}\quad,
\end{equation}
where $\kappa^2=8\pi G$ and $\mathcal{L}^{(m)}$ is the Lagrangian
density for the matter field. We consider a homogeneous and
isotropic background universe described by the flat
Friedmann-Robertson-Walker(FRW) metric
\begin{equation}
ds^2=-dt^2+a^2dx^2\quad,
\end{equation}
The background dynamics of the universe in $f(R)$ gravity is given by\cite{frreview}
\begin{equation}
H^2=\frac{FR-f}{6F}-H\frac{\dot{F}}{F}+\frac{\kappa^2}{3F}\rho\label{field}\quad.
\end{equation}
where $\rho=\rho_m+\rho_r$, $F=\frac{\partial f(R)}{\partial R}$. If
we convert the derivative from the cosmic time $t$ to $x=\ln a$ and
further take the derivative of the above equation, we obtain
\begin{equation}
\frac{d^2}{dx^2}F+(\frac{1}{2}\frac{d\ln E}{dx}-1)\frac{dF}{dx}+\frac{d\ln E}{dx}F=\frac{\kappa^2}{3E}\frac{d\rho}{dx}\quad ,\label{Ffield}
\end{equation}
where
\begin{equation}
\begin{split}
E&\equiv\frac{H^2}{H_0^2}\quad ,\\
R&\equiv3(\frac{d E}{dx}+4E)\quad ,\\
\frac{d\rho}{dx}&=-3(\rho+p)\quad.\label{Rdef}
\end{split}
\end{equation}
For convenience, we take the energy density $\rho$ and the scalar curvature $R$ in Eq.(\ref{Ffield}) in the unit of $H_0^2$ and we also set $\kappa^2=1$ in our analysis. In order to get a viable $f(R)$ model with a reasonable expansion history of the universe allowed by observations, we can parameterize $E(x)$ in Eq.(\ref{Ffield}) as the standard model in Einstein's gravity with an effective dark energy equation of state(EoS) $w$\cite{Song}\cite{Pogosian}
\begin{equation}
E(x)=\Omega_r^0e^{-4x}+\Omega_m^0e^{-3x}+\Omega_d^0e^{-3\int_0^x(1+w)dx}\quad,\label{Ex}
\end{equation}
where
\begin{equation}
\begin{split}
\Omega_m^0&\equiv\frac{\kappa^2\rho_m^0}{3H_0^2}\quad,\\
\Omega_d^0&\equiv\frac{\kappa^2\rho_d^0}{3H_0^2}\quad,\\
\Omega_r^0&\equiv\frac{\kappa^2\rho_r^0}{3H_0^2}\quad.\label{defination}
\end{split}
\end{equation}

After specifying the expansion history of the universe $E(x)$, Eq.(\ref{Ffield}) becomes a second order differential equation of $F(x)$. If we can find the solution of Eq.(\ref{Ffield}), we then obtain the explicit form of $f(R)$ correspondingly.  We find that it is more convenient to use the quantity $G=F-1$ instead of $F$, so that Eq.(\ref{Ffield}) can be changed into
\begin{equation}
\begin{split}
&\frac{d^2G}{dx^2}+(\frac{1}{2}\frac{d\ln E}{dx}-1)\frac{dG}{dx}+\frac{d\ln E}{dx}G\\
&=\frac{3(1+w)\Omega_d^0}{E}e^{-3\int_0^x(1+w)dx}\quad.\label{Gfield}
\end{split}
\end{equation}
If we want to mimic the exact $\Lambda$CDM expansion history of the universe $w=-1$, Eq.(\ref{Gfield}) is a homogenous equation.
We will focus on this case hereafter and show that the solution of the above differential equation will  directly lead to the real-valued $f(R)$ form which gives rise to the cosmological evolution as that of the $\Lambda$CDM model.

Eq.(\ref{Gfield}) does not have an analytical solution to describe the universe from the radiation dominated epoch to the late time acceleration with the full expression of Eq.(\ref{Ex}) . However, Eq.(\ref{Gfield}) does have analytical solutions in different epochs in the evolution of the universe. Let's first concentrate on  the early evolution of the universe from the radiation dominated epoch to the matter dominated epoch. In this case, $E$ can be taken as
\begin{equation}
E_1\sim\Omega_m^0e^{-3x}+\Omega_r^0e^{-4x}\nonumber\quad,
\end{equation}
and the general solution of Eq.(\ref{Gfield}) has the form
\begin{equation}
\begin{split}
G_1(x)&= C_1 G^{22}_{22}\left(
                 \begin{array}{cc}
                   n_- & n_+ \\
                   -1 & 4 \\
                 \end{array}|-\frac{\Omega_m^0}{\Omega_r^0}e^x
               \right)\\
&+D_1e^{4x}{_2F_1}[m_-,m_+;6;-\frac{\Omega_m^0}{\Omega_r^0}e^x]\label{RMsolution}
\end{split}
\end{equation}
where $G^{22}_{22}$ is the Meijer G function, ${_2F_1}$ is the Gaussian hypergeometric function and $C_1$, $D_1$ are arbitrary constants which can be determined by boundary conditions.

The indexes in the solutions are
\begin{eqnarray}
m_+&=&\frac{11+\sqrt{73}}{4}\nonumber \quad,\\
m_-&=&\frac{11-\sqrt{73}}{4}\nonumber \quad,\\
n_+&=&\frac{9+\sqrt{73}}{4}\nonumber \quad,\\
n_-&=&\frac{9-\sqrt{73}}{4}\nonumber \quad .
\end{eqnarray}
The viable $f(R)$ models should be of the ``chameleon" type \cite{Mota}\cite{Khoury} which provides a mechanism to pass the local test. However, the first term of Eq.(\ref{RMsolution}) is divergent when $x$ goes to minus infinity. Thus the requirement $\lim_{x\rightarrow -\infty} G(x)=0$ puts the condition $C_1=0$, so that $G_1(x)$ turns out to be
\begin{equation}
G_1(x)= D_1e^{4x}{_2F_1}[m_-,m_+;6;-\frac{\Omega_m^0}{\Omega_r^0}e^x].
\end{equation}
We can obtain the explicit form for the $f(R)$ model to describe the early universe by doing the integration
\begin{equation}
f(R)=R+\int G(x)\frac{dR}{dx}dx\quad\label{generalfr}\quad,
\end{equation}
where the scalar curvature $R$ can be written as
\begin{equation}
R\rightarrow3\Omega_m^0e^{-3x}\label{Rtox}\quad,
\end{equation}
which only contains the component of matter because the radiation does not have any contribution to the scalar curvature $R$ since its energy momentum tensor is traceless.
Eq.(\ref{Rtox}) is valid during the expansion history of the universe from the radiation dominated epoch to the deep matter dominated epoch.  The term $\frac{dR}{dx}$ in Eq.(\ref{generalfr}) thus can be expressed as
\begin{equation}
\frac{dR}{dx}=-9\Omega_m^0 e^{-3x}\nonumber \quad ,
\end{equation}
where $x$, in turn, can be presented in terms of $R$
\begin{equation}
x(R)=-\frac{1}{3}\ln \left( \frac{R}{3\Omega_m^0} \right)\quad.\label{xtoR}
\end{equation}
When $x$ goes to minus infinity $x\rightarrow -\infty$, $R$ goes to infinity $R\rightarrow +\infty$. Combining the above equations, we obtain the explicit expression for $f(R)$ as
\begin{equation}
\begin{split}
f_1(R)&=R-\frac{45\Omega_r^0D_1}{(m_+-1)(m_--1)}\\
&+\frac{45\Omega_r^0D_1}{(m_+-1)(m_--1)}\times\\
&{_2F_1}\left [m_--1,m_+-1;5;-\frac{\Omega_m^0}{\Omega_r^0}\left
(\frac{3\Omega_m^0}{R}\right )^{1/3}\right]\\
 &+C_I\quad,
\end{split}
\label{radiation2}
\end{equation}
where $f_1(R)$ and $R$ are in the unit of $H_0^2$. The
additional constant $C_I$ arises from the indefinite integral of
Eq.(\ref{generalfr}). When $D_1=0$, $f_1(R)$ should go back to the
standard Einstein's gravity, namely $f_1(R)=R$, such that the
constant of integration $C_I$  in Eq.(\ref{radiation2})
vanishes. Eq.(\ref{radiation2}) is
mathematically well-defined for all positive values of the scalar
curvature $R>0$ because the hypergeometric function
${_2F_1}[a,b;c;z]$ has the integral representation on the real axis
when $b>0$ and $c>0$
\begin{equation}
{_2F_1}[a,b;c;z]=\frac{\Gamma(c)}{\Gamma(b)\Gamma(c-b)}\times\int_0^{1}t^{b-1}(1-t)^{c-b-1}(1-zt)^{-a}dt\quad,\label{defhypergeometric}
\end{equation}
where $\Gamma$ is the Euler gamma function. The above expression is well-defined in the range $-\infty<z<1$ and the resulting value of ${_2F_1}[a,b;c;z]$ is also a real value. In order to present the $f(R)$ in SI units, we can insert Eq.(\ref{defination}) and then we obtain
\begin{equation}
\begin{split}
f_1(R)&= R-\frac{15\kappa^2\rho_r^0D_1}{(m_+-1)(m_--1)}\\
&+\frac{15\kappa^2\rho_r^0D_1}{(m_+-1)(m_--1)}\times\\
&{_2F_1}\left [m_--1,m_+-1;5;-\frac{\rho_m^0}{\rho_r^0}\left (\frac{R_0-4\Lambda}{R}\right )^{1/3}\right]\quad\label{radiation} .
\end{split}
\end{equation}
where $\rho_r^0$ and  $\rho_m^0$ are the energy density of radiation
and matter at today respectively. $R_0$ is the scalar curvature
$R_0=\kappa^2\rho_m^0+4\Lambda$.  Eq.(\ref{radiation}) shows that we
have the well-defined real analytical function for $f(R)$ gravity
that can exactly reproduce the same background expansion as that of
the $\Lambda$CDM model from the radiation dominated epoch to the
matter dominated epoch.

At the very early time of the universe, when the universe is dominated by the radiation and the curvature is very high $R\gg4\Lambda$, we have
\begin{equation}
\begin{split}
&{_2F_1}\left [m_--1,m_+-1;5;-\frac{\rho_m^0}{\rho_r^0}\left (\frac{R_0-4\Lambda}{R}\right )^{1/3}\right]\\
\approx&1-\frac{1}{5}(m_--1)(m_+-1)\frac{\rho_m^0}{\rho_r^0}\left (\frac{R_0-4\Lambda}{R}\right )^{1/3}\quad.
\end{split}
\label{expansion}
\end{equation}
Eq.(\ref{radiation}), thus, reduces to
\begin{equation}
f_1(R)\sim R-D_13\kappa^2\rho_m^0\left (\frac{\kappa^2\rho_m^0}{R}\right )^{1/3}\label{radiationfr}
\end{equation}
The above expression is just the $f(R)$ model which can exactly recover the same radiation dominated expansion history of the universe as that of the LCDM model. Eq.(\ref{radiationfr}) is  consistent with the result obtained from Eq.(\ref{Gfield}) by setting
$E_1\sim\Omega_r^0e^{-4x}$ directly.

When the universe evolved from the radiation dominated epoch to the matter dominated era, we have
\begin{equation}
z\equiv\frac{\rho_m^0}{\rho_r^0}\left (\frac{R_0-4\Lambda}{R}\right )^{1/3}\sim\frac{\rho_m^0}{\rho_r^0}e^{x}>>1\quad,
\end{equation}
since $e^{x}\sim0.01$ and $\Omega_r^0<<\Omega_m^0$. The term ${_2F_1}\left [m_--1,m_+-1;5;-z\right]$ in Eq.(\ref{radiation}), can be expanded around $z=+\infty$ as,
\begin{equation}
{_2F_1}\left [m_--1,m_+-1;5;-z\right]\approx \frac{24\Gamma(\frac{\sqrt{73}}{2})}{\Gamma(\frac{7+\sqrt{73}}{4})\Gamma(\frac{13+\sqrt{73}}{4})}z^{\frac{-7+\sqrt{73}}{4}}
\end{equation}
and Eq.(\ref{radiation}), thus, reduces to
\begin{equation}
f_1(R)\sim R-\xi_1\left (\frac{\Lambda}{R}\right )^{p_+-1}\label{axmatterfr}
\end{equation}
where $p_+=\frac{5+\sqrt{73}}{12}$ and $\xi_1$ is given by
\begin{equation}
\xi_1=\frac{240D_1\Gamma(\frac{\sqrt{73}}{2})}{\Gamma(\frac{7+\sqrt{73}}{4})\Gamma(\frac{13+\sqrt{73}}{4})}
\frac{\kappa^2\rho_r^0\left(R_0-4\Lambda\right)^{p_+-1} }{\Lambda^{p_+-1}}\left(\frac{\rho_m^0}{\rho_r^0}\right)^{3(p_+-1)}\quad.
\end{equation}
Eq.(\ref{axmatterfr}) is  consistent  with the result obtained from Eq.(\ref{Gfield}) by setting
$E_1\sim\Omega_m^0e^{-3x}$ directly, which represents the $f(R)$ model that can mimic the LCDM expansion history of the universe in the matter dominated phase.

\section{The transition from the  matter dominated epoch to the late time acceleration era\label{MA}}
Next we turn to investigate the most interesting case that the $f(R)$ model can mimic the evolution of the universe from the matter dominated epoch to the late time acceleration. In this case, $E$ can be taken as
\begin{equation}
E_2=\Omega_m^0e^{-3x}+\Omega_d^0\nonumber\quad,
\end{equation}
where $\Omega_d^0$ is a constant. The scalar curvature can be presented as
\begin{equation}
R=3\Omega_m^0e^{-3x}+12\Omega_d^0\quad.\label{Rscalar}
\end{equation}
The general solution of Eq.(\ref{Gfield}) gives
\begin{equation}
\begin{split}
G_2(x)&= C_2(e^{3x})^{p_-}{_2F_1}[q_-,p_-;r_-;-e^{3x}\frac{\Omega_d^0}{\Omega_m^0}]\\
&+D_2(e^{3x})^{p_+}{_2F_1}[q_+,p_+;r_+;-e^{3x}\frac{\Omega_d^0}{\Omega_m^0}] \quad\label{mG},
\end{split}
\end{equation}
where the indexes are
\begin{eqnarray}
q_+&=&\frac{1+\sqrt{73}}{12}\nonumber \quad,\\
q_-&=&\frac{1-\sqrt{73}}{12}\nonumber \quad,\\
r_+&=&1+\frac{\sqrt{73}}{6}\nonumber \quad,\\
r_-&=&1-\frac{\sqrt{73}}{6}\nonumber\quad,\\
p_+&=&\frac{5+\sqrt{73}}{12}\nonumber \quad,\\
p_-&=&\frac{5-\sqrt{73}}{12}\nonumber \quad.
\end{eqnarray}


After doing the integration, we can get the explicit expression for
$f(R)$ from Eq.(\ref{generalfr})
\begin{equation}
f_2(R)=R-12\Omega_d^0\label{matter2}-\epsilon_+-\epsilon_-+C_I\quad
,
\end{equation}
where
\begin{equation}
\begin{split}
\epsilon_+&=\frac{D_2(3\Omega_m^0)}{p_+-1}\left (\frac{3\Omega_m^0}{R-12\Omega_d^0}\right )^{p_+-1}\times\\
&{_2F_1}\left[q_+,p_+-1;r_+;-\frac{3\Omega_d^0}{R-12\Omega_d^0}\right
]\nonumber\quad.
\end{split}
\end{equation}
\begin{equation}
\begin{split}
\epsilon_-&=\frac{C_2(3\Omega_m^0)}{p_--1}\left (\frac{3\Omega_m^0}{R-12\Omega_d^0}\right )^{p_--1}\times\\
&{_2F_1}\left[q_-,p_--1;r_-;-\frac{3\Omega_d^0}{R-12\Omega_d^0}\right
]\nonumber\quad.
\end{split}
\end{equation}
The constant of integration $C_I$ should be chosen as
$6\Omega_d^0$ such that when $C_2=D_2=0$, $f_2(R)$ goes back to the
standard Einstein's gravity with cosmological constant, namely
$f_2(R)=R-6\Omega_d^0$.

Inserting Eq.(\ref{defination}), we obtain
\begin{equation}
\begin{split}
f_2(R)&=R-2\Lambda\\
&-\varpi_1\left (\frac{\Lambda}{R-4\Lambda}\right )^{p_+-1}{_2F_1}\left[q_+,p_+-1;r_+;-\frac{\Lambda}{R-4\Lambda}\right ]\\
&-\varpi_2\left (\frac{\Lambda}{R-4\Lambda}\right )^{p_--1}{_2F_1}\left[q_-,p_--1;r_-;-\frac{\Lambda}{R-4\Lambda}\right ]
\label{viable}
\end{split}
\end{equation}
where $\varpi_1= D_2(R_0-4\Lambda)^{p_+}/(p_+-1)/\Lambda^{p_+-1}$ and $\varpi_2= C_2(R_0-4\Lambda)^{p_-}/(p_--1)/\Lambda^{p_--1}$. The constant parameter $\Lambda $ is defined as $ \Lambda\equiv\kappa^2\rho_d$ and $\rho_d$ is the effective energy density of dark energy.  When $\varpi_1=\varpi_2=0$, $\Lambda $ is just the cosmological constant. Noting the fact that $p_--1<0$, when $R\rightarrow+\infty$, the last term of Eq.(\ref{viable}) becomes divergent. This is not allowed for the ``chameleon" type  solution, so that we need to set $\varpi_2=0$. Therefore the solution has the form
\begin{equation}
\begin{split}
f_2(R)&=R-2\Lambda\\
&-\varpi_1\left (\frac{\Lambda}{R-4\Lambda}\right )^{p_+-1}{_2F_1}\left[q_+,p_+-1;r_+;-\frac{\Lambda}{R-4\Lambda}\right ].
\end{split}
\label{viable2}
\end{equation}
From the integral representation of the hypergeometric function Eq.(\ref{defhypergeometric}), it is clear that Eq.(\ref{viable2}) is mathematically well-defined in the range $R> 4\Lambda$.  Eq.(\ref{viable2}) is a real function in the physical range from the matter dominated epoch to the future expansion of the universe.

In the matter dominated phase $R>>4\Lambda$, ${_2F_1}$ goes back to the unity and Eq.(\ref{viable2}) can be approximated as
\begin{equation}
f_2(R)\sim R-\varpi_1\left (\frac{\Lambda}{R}\right )^{p_+-1}\quad,
\end{equation}
which is consistent with Eq.(\ref{axmatterfr}).
By comparing the coefficient of the above equation with Eq.(\ref{axmatterfr}), we can obtain the relation between $D_1$ and $D_2$ as
\begin{equation}
D_2=D_1\frac{240(p_+-1)\Gamma(\frac{\sqrt{73}}{2})}{\Gamma(\frac{7+\sqrt{73}}{4})\Gamma(\frac{13+\sqrt{73}}{4})}
 \left(\frac{\rho_m^0}{\rho_r^0}\right)^{3p_+-4}
\end{equation}

Eq.(\ref{Gfield}) is our starting point to find the analytic expression for the $f(R)$ model to mimic the $\Lambda$CDM cosmology. From Eq.(\ref{Rscalar}), we can see clearly that the scalar curvature $R$  obtained from Eq.(\ref{Gfield}) can be constrained automatically in the physical range $R>4\Lambda$. Our starting point is different from that in ~\cite{dobado}~\cite{Dunsby}~\cite{Nojiri}, where they got their solution by solving the differential equation ~\cite{dobado}~\cite{Dunsby}~\cite{solution}~\cite{Nojiri}
\begin{equation}
\begin{split}
&3(R-3\Lambda)(R-4\Lambda)\frac{d^2}{dR^2}f(R)\\
=&(\frac{R}{2}-3\Lambda)\frac{d}{dR}f(R)+\frac{1}{2}f(R)+4\Lambda-R\label{hyeq}
\end{split}
\end{equation}
This equation was derived from Eq.(\ref{field}). In Eq.(\ref{hyeq}), $R$ can be chosen as any value on the real axes, so that not all solutions of Eq.(\ref{hyeq}) are physical. Thus, we need to carefully analyze the solutions of Eq.(\ref{hyeq}) .

A particular solution for Eq.(\ref{hyeq}) is
\begin{equation}
f_p(R)=R-2\Lambda\quad.
\end{equation}
Therefore, we only focus on the homogenous solutions of Eq.({\ref{hyeq}) hereafter since $f(R)=f_p(R)-f_h(R)$.
The homogenous part of Eq.({\ref{hyeq}) is a standard hypergeometric equation which may have at most $24$ solutions in the complex plane around three different singular points($R=\infty,3\Lambda,4\Lambda$)~\cite{hysolution}. However, if we focus on real solutions, Eq.({\ref{hyeq}) may have at most $32$ solutions on the real axes around four different singular points $R=-\infty,3\Lambda,4\Lambda,+\infty$ . We will extensively discuss all of these solutions in the following.

The solution around $+\infty$ reads,
\begin{equation}
\begin{split}
&f_h(R)\\
=&\varpi_1\left (\frac{\Lambda}{R-4\Lambda}\right )^{p_+-1}{_2F_1}\left[q_+,p_+-1;r_+;-\frac{\Lambda}{R-4\Lambda}\right ]\\
+&\varpi_2\left (\frac{\Lambda}{R-4\Lambda}\right )^{p_--1}{_2F_1}\left[q_-,p_--1;r_-;-\frac{\Lambda}{R-4\Lambda}\right ]
\quad,\label{possitive}
\end{split}
\end{equation}
The above expression is just Eq.(\ref{viable}) which is the physical solution of Eq.(\ref{hyeq}).
We can see clearly that when $R\rightarrow +\infty$, $f_h(R)$ is well-defined on the real axes and actually $f_h(R)$ is   a real function for the whole range of $R>4\Lambda$ as discussed previously. $R=4\Lambda$ is a finite point because $\lim_{R\rightarrow4\Lambda}f_h(R)$ is a finite value. However, the derivatives $\frac{d}{dR}f_h(R)$ of all the terms in the above expression are divergent at $R=4\Lambda$.

The solution around $-\infty$ reads
\begin{equation}
\begin{split}
&f_h(R)\\
=&\varpi_1\left (\frac{-\Lambda}{R-3\Lambda}\right )^{p_+-1}{_2F_1}\left[p_+-1,r_+-q_+;r_+;\frac{\Lambda}{R-3\Lambda}\right]\\
+&\varpi_2\left (\frac{-\Lambda}{R-3\Lambda}\right )^{p_--1}{_2F_1}\left[p_--1,r_--q_-;r_-;\frac{\Lambda}{R-3\Lambda}\right]
\label{negative}
\end{split}
\end{equation}
This expression was obtained in~\cite{dobado}. This solution is well-defined when $R\rightarrow-\infty$. It is a real function when $R<3\Lambda$. However, it becomes complex when $R>3\Lambda$. $R=3\Lambda$ is a finite point. $f_h(R)$ and its derivative $\frac{d}{dR}f_h(R)$ are well defined at $R=3\Lambda$. Clearly Eq.(\ref{negative}) is not a physical solution, it can not satisfy Eq.(\ref{Gfield}).

The solution around $3\Lambda$ reads,
\begin{equation}
\begin{split}
f_h(R)&=\varpi_1{_2F_1}\left[\alpha_+,\alpha_-;-\frac{1}{2};\frac{R}{\Lambda}-3\right]\\
&+\varpi_2\left (\frac{R}{\Lambda}-3\right )^{3/2}{_2F_1}\left[\beta_+,\beta_-;\frac{5}{2};\frac{R}{\Lambda}-3\right]\quad,\\
\end{split}\label{3lambda}
\end{equation}
where $\alpha_{\pm}=(-7\pm\sqrt{73})/12$ and $\beta_{\pm}=(11\pm\sqrt{73})/12$. The above solution was derived in ~\cite{Dunsby}~\cite{Nojiri}. $f_h(R)$ and its derivative $\frac{d}{dR}f_h(R)$ are well-defined on the finite singular point $R=3\Lambda$.  When $R>4\Lambda$ or $R<3\Lambda$, the second term of Eq.(\ref{3lambda}) becomes complex. However, in contrast to what was claimed in \cite{Dunsby}, when $R>4\Lambda$ the homogenous part of Eq.(\ref{hyeq}) do have the real analytical solution, namely Eq.(\ref{possitive}).

The solution around $4\Lambda$ reads,
\begin{equation}
\begin{split}
f_h(R)&=\varpi_1{_2F_1}\left[p_+-1,p_--1;\frac{1}{3};-(\frac{R}{\Lambda}-4)\right]\\
&+\varpi_2\left (\frac{R}{\Lambda}-4\right )^{2/3}{_2F_1}\left[q_+,q_-;\frac{5}{3};-(\frac{R}{\Lambda}-4)\right]\quad\\
\end{split}\label{lambda4}
\end{equation}
Clearly the above solution is well-defined on the finite singular point $R=4\Lambda$ and the solution is valid for $R\geq4\Lambda$. It is a solution in the physical range and satisfies Eq.(\ref{Gfield}).
Eq.(\ref{lambda4}) is equivalent to Eq.(\ref{possitive}) since they are defined in the same range.  Eq.(\ref{lambda4}) is simply a new linear combination of the solutions in Eq.(\ref{possitive}). However, Eq.(\ref{lambda4}) has different behaviors at singular points if compared with Eq.(\ref{possitive}) owing to the different linear combination of hypergeometric functions.
For instance, the derivatives $\frac{d}{dR}f_h(R)$ of all the terms in Eq.(\ref{possitive}) are divergent at $R=4\Lambda$ while in Eq.(\ref{lambda4}), the derivative of the first term ${_2F_1}\left[p_+-1,p_--1;\frac{1}{3};-(\frac{R}{\Lambda}-4)\right]$ is well-defined at $R=4\Lambda$.
Furthermore, Eq.(\ref{lambda4}) apparently does not have the ``chameleon" property because both terms in Eq.(\ref{lambda4}) are divergent when $R$ goes to infinity. However, their linear combination, namely, the term in Eq.(\ref{viable2}) is convergent when $R\rightarrow+\infty$. Therefore, even for equivalent results, we should carefully choose the proper expressions according to boundary conditions at singular points.

Using the Euler transformation and Pfaff transformation, from Eqs.(\ref{possitive},\ref{negative},\ref{3lambda},\ref{lambda4}), we can find out all the $8\times4=32$ real solutions for Eq.(\ref{hyeq}).  Eqs.(\ref{possitive},\ref{negative},\ref{3lambda},\ref{lambda4}) complete the different behaviors at different singular points for the solutions of Eq.(\ref{hyeq}). Although Eqs.(\ref{possitive},\ref{negative},\ref{3lambda},\ref{lambda4}) are substantially different on the real axes, when extended to the complex plane, Eqs.(\ref{possitive},\ref{negative},\ref{3lambda},\ref{lambda4}) are equivalent to each other because they can be related by connection formulas. However, for different expressions, they have different behaviors at the singular points. We should be very careful to choose the proper expressions according to different boundary conditions. The physical solutions for $f(R)$ models to describe the universe should be well-defined in the range $R>4\Lambda$ and possess the ``chameleon" property $\lim_{R\rightarrow +\infty} f_h(R)=0,\quad \lim_{R\rightarrow +\infty} \frac{d}{dR}f_h(R)=0$. The only physical solution is the ``chameleon" part of Eq.(\ref{possitive})($\varpi_2=0$), namely  Eq.(\ref{viable2}). The result obtained from Eq.(\ref{hyeq}) has bigger range in the solutions than that obtained from Eq.(\ref{Gfield}). We need to pick the physical solution very carefully. It is more convenient to start from Eq.(\ref{Gfield}) to find physical solutions.

Having the well-defined analytical expression for $f(R)$ model, we will further discuss its viability. Combining $f_1(R)$ and $f_2(R)$, it covers the entire expansion history of the universe from the radiation dominated epoch to the future expansion. However, we need to point out that the model $f_2(R)$, itself, is independently valid for the entire expansion history of the universe since in the very early time of the universe, we have $f_1(R)\sim f_2(R)\sim R$.  Therefore, in the next discussion, we will only focus on $f_2(R)$.

In order to evade the instabilities of the $f(R)$ model, we require that $F>0$ and $f_{RR}>0$~\cite{frreview}~\cite{Ignacy}. From Fig.(\ref{Fx}) and Fig.(\ref{fRR}), we can see that when $D_2<0$, in the past expansion of the universe $x<0$, we have $F_2(R)>0,f_{2RR}(R)>0$ so that the model $f_2(R)$ is viable in the past. However, for the future expansion of the universe, the condition $D_2<0$ can not guarantee $F_2(x)$ to be always positive.
The zero-crossing behavior will lead to singularities in the conformal transformation\cite{sigular} and the negative $F$ will lead to the imaginary mass of particles $\tilde{m}=m/\sqrt{F}$ in the Einstein frame ~\cite{frreview}~\cite{Hefr}. Furthermore, we can see  from Fig.(\ref{Fx}) and Fig.(\ref{fRR}) that, when $x\rightarrow+\infty$ or $R\rightarrow4\Lambda$, the derivative of $f(R)$, namely, $F(R)=\frac{d}{dR}f(R)$ and $f_{RR}(R)=\frac{d}{dR}F(R)$ are divergent. Although the expression of $f_2(R)$ has such weakness in the future evolution of the universe in the Einstein frame, $f_2(R)$ does not have the future singularity in the Lagrangian in the Jordan frame because $f_2(R)$ is finite at $R=4\Lambda$. The future point happens at
\begin{figure}
\includegraphics[width=2.5in,height=2.5in]{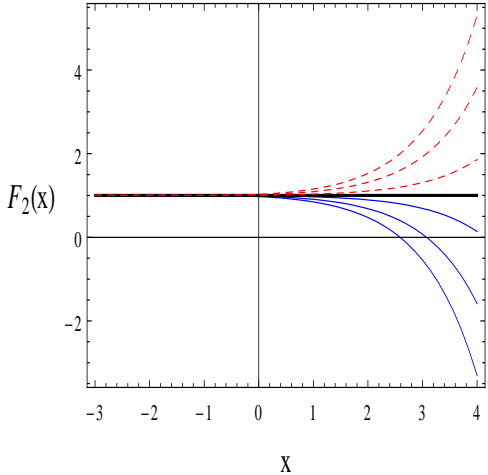}
\caption{ The red dashed lines from top to bottom represent
$D_2=0.05,0.03,0.01$, respectively. The thick black line represents the LCDM model with
$D_2=0$. The blue solid lines from top to bottom represent $D_2=-0.01,-0.03,-0.05$, respectively.
The models in red dashed lines are ruled out due to the instabilities in
the high curvature region. However, the scalar field $F(x)$ in blue
solid lines will become negative in the future expansion of the
universe. }\label{Fx}
\end{figure}
\begin{figure}
\includegraphics[width=2.7in,height=2.5in]{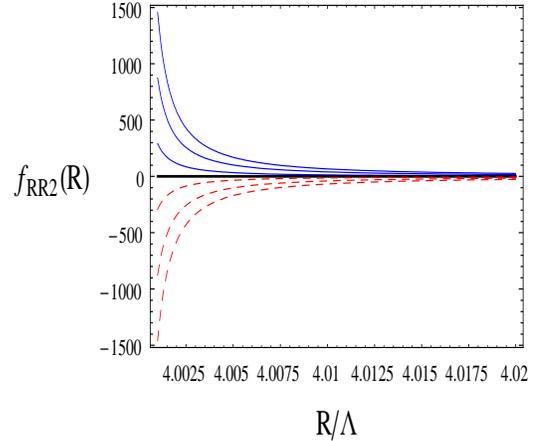}
\caption{The scalar field $f_{RR}$ will be divergent in the
future. Lines from top to bottom correspond to $D_2=-0.05,-0.03,-0.01,0,0.01,0.03,0.05$, respectively.
 }\label{fRR}
\end{figure}
\begin{equation}
\lim_{x\rightarrow+\infty}R=4\Lambda\quad,
\end{equation}
and $f_2(R)$ is finite at $R=4\Lambda$
\begin{equation}
\begin{split}
&\lim_{R\rightarrow4\Lambda}f_2(R)\\
=&2\Lambda-\varpi_1\frac{4(-511+79\sqrt{73})\Gamma(2/3)\Gamma(-r_-)}{(-5+\sqrt{73})(-1+\sqrt{73})(7+\sqrt{73})\Gamma(-p_-)\Gamma(q_+)}\\
\approx&2\Lambda-1.256\varpi_1\quad,
\end{split}
\end{equation}
when $\varpi<0$, we can find that $f_2(4\Lambda)>2\Lambda$.
\section{conclusions\label{conclusions}}
In summary, in this work we have constructed an $f(R)$ gravity model that mimics the $\Lambda$CDM universe expansion in both the early and late epochs. We found that there exists a real-valued function for the Ricci scalar in terms of hypergeometric functions which can give rise to the particular cosmological evolution of the $\Lambda$CDM model. Although the constructed $f(R)$ model has weakness in describing the future expansion of the universe in the Einstein frame, in the Jordan frame it is viable to describe the past evolution of the universe and it does not have the future singularity in the Lagrangian.

The fact that the Lagrangian is well-defined demonstrates that this family of $f(R)$
models are no longer just simply phenomenological models, but the field equations instead can be deduced from the principle of least action. Furthermore, when $w_1 \neq 0$, the constant
$\Lambda$ in Eq.(\ref{viable2}) cannot be explained as the energy density of the vacuum and the
model does not suffer the cosmological constant problem even though it has the same background expansion of the universe as the $\Lambda$CDM model. For the background
evolution of the universe, we cannot distinguish these $f(R)$ models from general relativity. However, we can distinguish them at the cosmological
perturbation level since the $f(R)$ gravity introduces an
extra scalar degree of freedom which has significant impact on perturbation equations. The constraints from observations at linear
perturbation level for the $f(R)$ model have been presented in our companion work
~\cite{He}. However, it would be more interesting to investigate the nonlinear behavior in f(R)
gravity using N-body simulations. The analytical functional form of $f(R)$ plays a vital role in
 N-body simulations, which is a subject of our future work.

\emph{Acknowledgments
J.H.He acknowledges the Financial
support of MIUR through PRIN 2008 and ASI through
contract Euclid-NIS I/039/10/0. The work of
B.Wang was partially supported by NNSF of China
under grant 10878001 and the National Basic
Research Program of China under grant
2010CB833000.}

\end{document}